\documentclass[prd,twocolumn,showpacs,floatfix,amsmath,nofootinbib,amssymb,floatfix]{revtex4}
\usepackage{graphicx,color,dcolumn,booktabs,bm,multirow}
\usepackage{longtable,lscape}
\usepackage{txfonts}
 \usepackage{enumitem}
\usepackage{overpic}
\usepackage{amssymb}
\usepackage{indentfirst}
\usepackage{feynmf}   
\usepackage{slashed}  
\usepackage{cases}
\usepackage{color}
\usepackage{multirow}
\usepackage{epstopdf}
\usepackage{graphicx,color,dcolumn,booktabs,bm}

\usepackage[colorlinks,
            citecolor=blue,
            anchorcolor=red,
            menucolor=red,
            linkcolor=red,
            filecolor=red,
            runcolor=red,
            urlcolor=blue,
            frenchlinks=red]{hyperref}
\usepackage{amsmath}
\usepackage{epsfig}
\usepackage{graphicx}
\usepackage{txfonts}

\begin{document}
\title{Possible charmed-strange molecular pentaquarks in quark delocalization color screening model }

\author{Xuejie Liu$^1$}\email[E-mail: ]{1830592517@qq.com}
\author{Yue Tan$^{2}$}\email[E-mail:]{tanyue@ycit.edu.cn}
\author{Xiaoyun Chen$^{5}$}\email[E-mail:]{xychen@jit.edu.cn}
\author{Dianyong Chen$^{1,3}$\footnote{Corresponding author}}\email[E-mail:]{chendy@seu.edu.cn}
\author{Hongxia Huang$^4$}\email[E-mail:]{hxhuang@njnu.edu.cn}
\author{Jialun Ping$^4$}\email[E-mail: ]{jlping@njnu.edu.cn}
\affiliation{$^1$School of Physics, Southeast University, Nanjing 210094, P. R. China}
\affiliation{$^2$School of Mathematics and Physics, Yancheng Institute of Technology, Yancheng, 224051,  P. R. China}
\affiliation{$^3$Lanzhou Center for Theoretical Physics, Lanzhou University, Lanzhou 730000, P. R. China}
\affiliation{$^4$Department of Physics, Nanjing Normal University, Nanjing 210023, P. R. China}
\affiliation{$^5$College of Science, Jinling Institute of Technology, Nanjing 211169, P. R. China}

\begin{abstract}
Inspired by the states $T_{c\bar{s}0}^{a}(2900)^{0}$ and $T_{c\bar{s}0}^{a}(2900)^{++}$ reported by the LHCb Collaboration, we carry out a systematical investigation of the charm-strange pentaquark system using resonance group method in the quark delocalization color screening model. The present results predict the existence of some bound states and resonance states with  support from the study of the mass spectrum and the decay properties. Both $\Sigma_{c}^{\ast}K^{\ast}$ with $I(J^{P})=\frac{1}{2}(\frac{5}{2}^{-})$ and $\Delta D_{s}^{\ast}$ with $I(J^{P})=\frac{3}{2}(\frac{5}{2}^{-})$ are bounded by channel coupling calculation. Moreover, the resonance state $\Sigma_{c}K^{\ast}$ with $I(J^{P})=\frac{1}{2}(\frac{1}{2}^{-})$ and $I(J^{P})=\frac{1}{2}(\frac{3}{2}^{-})$ are available in QDCSM, the masses and the total decay widths of which are ($R^{\prime}=3342\sim3346$ MeV, $\Gamma_{Total}=25.5$ MeV) and ($R^{\prime}=3333$ MeV, $\Gamma_{Total}=3.3$ MeV), respectively. In addition, the resonance state $\Delta D_{s}^{\ast}$ with $I(J^{P})=\frac{3}{2}(\frac{1}{2}^{-})$ is also obtained, the mass and the decay width of this state are 3343 MeV and 0.01 MeV, respectively. These predicted new exotic states may provide new ideas for experimental searches and we sincerely expect more experimental and theoretical research to verify and understand the charm-strange pentaquark states in the future.
\end{abstract}

\maketitle

\setcounter{totalnumber}{5}
\section{\label{sec:introduction}Introduction}

Since the discovery of $X(3872)$ in 2003, more and more exotic states have been discovered with the improvement of experimental equipment and techniques. It is hard to interpret these states simply as conventional baryons and mesons, and their inner structures are still under debate. It implies that these discovered states may be multiquark states. The notion of multiquark states had been proposed at the beginning of the construction of the quark model~\cite{Gell-Mann:1964ewy,Jaffe:1976yi}. Besides, the investigation of multiquark states has been of great help to an understanding of the non-perturbative QCD~\cite{Chen:2016qju, Swanson:2006st, Voloshin:2007dx, Chen:2016heh, Esposito:2016noz, Lebed:2016hpi, Guo:2017jvc}.

In the recent two decades, experimentally, an increasing number of charmed-strange-like states have been observed in addition to the charmonium-like state. As early as $2003$, the BABAR Collaboration reported a narrow peak $D_{s0}^{\ast}(2317)$  in the $D^{+}_{s}\pi$ invariant mass spectrum~\cite{BaBar:2003oey,BaBar:2004yux}. Later, the CLEO Collaboration~\cite{CLEO:2003ggt} confirmed the existence of this state and also reported another state $D_{s1}(2460)$. Besides, the existence of the $D_{s0}^{\ast}(2317)$ and $D_{s1}(2460)$ was confirmed by the Belle Collaboration~\cite{Belle:2003guh,Belle:2003kup} and BABAR Collaboration~\cite{BaBar:2003cdx,BaBar:2006eep}. Also, in 2018, BESIII Collaboration detected $D_{s0}^{\ast}(2317)$ by the observation of the process $e^{+}e^{-}\rightarrow D^{\ast+}_{s}D_{s0}^{\ast}(2317)+c.c$~\cite{BESIII:2017vdm}. In theory, since the mass positions of these two states are far from the theoretical predictions of the masses of charmed-strange mesons in the $J^{P}=0^{+}$ and $J^{P}=1^{+}$, so this inconsistency between the quark model expectations and experimental measurements makes these two states unlike conventional charmed-strange mesons. In addition to conventional mesons interpretation~\cite{Godfrey:2003kg,Rosner:2006jz,Godfrey:1985xj}, the molecular states ($DK$, $D^{\ast} K$)~\cite{Xiao:2016hoa,Barnes:2003dj,Navarra:2015iea,Kolomeitsev:2003ac,Hofmann:2003je,Guo:2006fu,Zhang:2006ix,Rosner:2006vc,Guo:2006rp,Liu:2022zbd}, tetraquark~\cite{Cheng:2003kg,Chen:2004dy,Kim:2005gt,Nielsen:2005ia,Terasaki:2005kc,Wang:2006uba} and a mixture of a $\bar{c}s$ meson and a tetraquark~\cite{Yang:2021tvc,Ortega:2016mms} interpretations had also been proposed, and the molecular state interpretation was more popular because the mass of these two states are just below the thresholds of $DK$ and $D^{\ast} K$.

The search for exotic states with charm and strangeness still goes on. In 2020, the LHCb Collaboration reported two fully open-flavor tetraquark candidates $X_{0}(2900)$ and $X_{1}(2900)$ in the invariant mass distribution of $D^{-}K^{+}$ of the channel $B^{\pm}\rightarrow D^{+}D^{-}K^{\pm}$~\cite{LHCb:2020bls,LHCb:2020pxc}. Since they are observed in the $D^{-}K^{+}$ spectrum, their valence quark contents are supposed to be $\bar{u}\bar{d}cs$. To reveal the nature of $X_{0}(2900)$ and $X_{1}(2900)$, there are also many theoretical explanations, such as the hadronic molecule states composed of $D^{\ast}\bar{K}^{\ast}$ or $D_{1}\bar{K}$~\cite{Wang:2020xyc,He:2020jna,Zhang:2020oze,Wang:2020prk}, compact tetraquark states~\cite{Liu:2020nil,Chen:2020aos,Huang:2020ptc,Molina:2020hde,Xue:2020vtq,Lu:2020qmp,Agaev:2020nrc,Mutuk:2020igv,Xiao:2020ltm,He:2020btl}, threshold effects~\cite{Liu:2020orv,Burns:2020epm} and so on.

Very recently, the LHCb Collaboration reported the observation of two new tetraquark candidates in the $B^{0}\rightarrow \bar{D}^{0} D_{s}^{+}\pi^{-}$ and $B^{+}\rightarrow D^{-}D^{+}_{s}\pi^{+}$ decays~\cite{Tcs(2900):exp}. They were named $T_{c\bar{s}}(2900)^{0}$ and $T_{c\bar{s}}(2900)^{++}$. From the decay channels, the least quark contents of these two states were obviously $c\bar{s}q\bar{q}$ (q=u, d), which are similar to the $X_{0}(2900)$ and $X_{1}(2900)$, and their quantum numbers are both $I(J^{P})=0(1^{+})$. Actually, there is a lot of work that has been done on open-flavor states~\cite{Ge:2022dsp,Ke:2022ocs,Wei:2022wtr,Chen:2022svh,Agaev:2022duz,He:2020jna,Cheng:2020nho,Albuquerque:2020ugi}.

Inspired by the experimental discovery of charmed-strange exotic states, such as $T_{c\bar{s}}(2900)^{0}$ and $T_{c\bar{s}}(2900)^{++}$, whether the physical world could have pentaquark states with charm and strangeness. To test this conjecture, in this work we replace $\bar{n}$ in $T_{c\bar{s}}(2900)$ by $nn$ quark pairs and this substitution is to some extent equivalent. From this point of view, the experimentally observed $T_{c\bar{s}}$ suggest that charmed-strange pentaquark states may exist. According to the quark component of $c\bar{s}nnn$ system,    since quark-antiquark pair is not identical particles, they cannot be annihilated in a vacuum, which means that the $c\bar{s}nnn$ states can be definitely charmed-strange pentaquark state once discovered. Similar related work had been done in theory. For example, in Ref.~\cite{Chen:2022svh}, the $\Lambda_{c}K^{(\ast)}$ and $\Sigma_{c}K^{(\ast)}$ interactions were investigated by adopting the OBE effective potentials and considering the $S-D$ wave mixing effects. The results predicted four possible charmed-strange molecular pentaquarks. In Ref.~\cite{An:2022vtg}, the authors systematically studied the mass spectrum and decay properties of the charmed-strange pentaquark system $c\bar{s}nnn$ and suggested experiments to search them in the b-hadron decays.

In this work, to determine the possibility of the existence of charmed-strange pentaquark states, we systematically investigate the $c\bar{s}nnn$ system in the quark delocalization color screening model.
Firstly, we calculate the effective potential between two hadrons in different quantum regimes to evaluate the properties of their interactions. Then to confirm the existence of bound states, we perform the bound calculation which takes into account channel coupling effects. Finally, based on the conservation of the quantum numbers and the limit of phase space, we investigate the possible strong decay channels of the charmed-strange pentaquark system to determine the existence of resonance states.

 The rest of this paper is organized as follows. In Sect.~\ref{mod}, the detail of the quark delocalization color screening model (QDCSM) is presented. The effective potential, the bound calculation, and the scattering phase shift calculation is given in Sect.~\ref{dis}, and the discussion and analysis of these results are also presented in Sect.~\ref{dis}. A brief summary is given in the last section.




\section{THE QUARK DELOCALIZATION COLOR SCREENING MODEL (QDCSM) \label{mod}}

The quark delocalization color screening model (QDCSM) is an extension of the native quark cluster model~\cite{DeRujula:1975qlm,Isgur:1979be,Isgur:1978wd,Isgur:1978xj} and was developed with aim of
addressing multiquark systems. The detail of QDCSM can be found in the Refs.~\cite{Wang:1992wi,Chen:2007qn,Chen:2011zzb,Wu:1996fm,Huang:2011kf}.
Here, the general form of the five-body complex Hamiltonian is given by
\begin{equation}
H = \sum_{i=1}^{5} \left(m_i+\frac{\boldsymbol{p}_i^2}{2m_i}\right)-T_{CM}+\sum_{j>i=1}^5V(r_{ij}),\\
\end{equation}
where the center-of-mass kinetic energy, $T_{CM}$ is subtracted without losing generality since we mainly focus on the internal relative motions of the multiquark system. The interplay is of two body potentials which includes color-confining, $V_{CON}$, one-gluon exchange, $V_{OGE}$, and Goldstone-boson exchange, $V_{\chi}$, respectively,
\begin{equation}
V(r_{ij}) = V_{CON}(r_{ij})+V_{OGE}(r_{ij})+V_{\chi}(r_{ij}).
\end{equation}

Note herein that the potential could contain central, spin-spin, spin-orbit, and tensor contributions; In this work, only the first two will be considered attending the goal of the present calculation and for clarity in our discussion.
The potential $V_{OGE}(r_{ij})$ can be written as
\begin{eqnarray*}
V_{OGE}(r_{ij}) &=& \frac{1}{4}\alpha_{ij} \boldsymbol{\lambda}^{c}_i \cdot\boldsymbol{\lambda}^{c}_j \\
&&\left[\frac{1}{r_{ij}}-\frac{\pi}{2}\delta(\boldsymbol{r}_{ij})\left(\frac{1}{m^2_i}+\frac{1}{m^2_j}
+\frac{4\boldsymbol{\sigma}_i\cdot\boldsymbol{\sigma}_j}{3m_im_j}\right)\right],
\end{eqnarray*}
where $m_{i}$ and $\boldsymbol{\sigma}$ are the quark mass and the Pauli matrices, respectively. The $\boldsymbol{\lambda^{c}}$ is SU(3) color matrix. The QCD-inspired effective scale-dependent strong coupling constant, $\alpha_{ij}$, offers a consistent description of mesons and baryons from the light to the heavy quark sector. It is associated with the quark flavor and determined by the mass difference of two hadrons.

Similary, the confining interaction $V_{CON}(r_{ij})$ can be expressed as
\begin{equation}
 V_{CON}(r_{ij}) =  -a_{c}\boldsymbol{\lambda^{c}_{i}\cdot\lambda^{c}_{j}}\left[f(r_{ij})+V_{0_{ij}}\right],
\end{equation}
where the $V_{0_{ij}}$ is the zero-point potential, which is determined by the mass shift of the absolute and experimental value of meson or baryon, and it is also related to the quark flavor. Moreover, the $f(r_{ij})$ can be written as
\begin{equation}
 f(r_{ij}) =  \left\{ \begin{array}{ll}r_{ij}^2 &\qquad \mbox{if }i,j\mbox{ occur in the same cluster}, \\
\frac{1 - e^{-\mu_{ij} r_{ij}^2} }{\mu_{ij}} & \qquad \mbox{if }i,j\mbox{ occur in different cluster}, \\
\end{array} \right.
\end{equation}
where the color screening parameter $\mu_{ij}$ is determined by fitting the deuteron properties, $NN$ and $NY$ scattering phase shifts~\cite{Chen:2011zzb,Ping:1993me,Wang:1998nk}., with $\mu_{qq}= 0.45$, $\mu_{qs}= 0.19$
and $\mu_{ss}= 0.08$, satisfying the relation $\mu_{qs}^{2}=\mu_{qq}\mu_{ss}$, where $q$ represents $u$ or $d$ quark. When extending to the heavy-quark case, we found that the dependence of the parameter $\mu_{cc}$ is not very significant in the calculation of the $P_{c}$ states~\cite{Huang:2015uda} by taking it from $10^{-4}$ to $10^{-2}\ \mathrm{fm}^{-2}$. The typical size of the multiquark system is several femtometres, thus the value of the $\mu_{ij} r^2$ is rather small, and in this case, the exponential function can be approximated to be
\begin{eqnarray}\label{muij}
  e^{-\mu_{ij}r_{ij}^{2}} &=& 1-\mu_{ij}r_{ij}^{2}+\mathcal{O}\left(\mu_{ij}^2 r_{ij}^4\right).
\end{eqnarray}
Accordingly, the confinement potential between two clusters is approximated to be
\begin{eqnarray}
  V_{CON}(r_{ij}) &=&  -a_{c}\boldsymbol{\mathbf{\lambda}}^c_{i}\cdot
\boldsymbol{\mathbf{
\lambda}}^c_{j}~\left(\frac{1-e^{-\mu_{ij}\mathbf{r}_{ij}^2}}{\mu_{ij}}+
V_{0_{ij}}\right) \nonumber \\
  ~ &\approx & -a_{c}\boldsymbol{\mathbf{\lambda}}^c_{i}\cdot
\boldsymbol{\mathbf{ \lambda}}^c_{j}~\left(r_{ij}^2+ V_{0_{ij}}\right),
\end{eqnarray}
which is the same as the expression of two quarks in the same cluster. Thus, when the value of the $\mu_{ij}$ is very small, the screened confinement will return to the quadratic form, which is why the results are insensitive to the value of $\mu_{cc}$. In the present work, we take $\mu_{cc}=0.01$. Then $\mu_{sc}$ and $\mu_{uc}$ are obtained by the relation $\mu_{sc}^{2}=\mu_{ss}\mu_{cc} $ and $\mu_{uc}^{2}=\mu_{uu}\mu_{cc}$, respectively. Besides, as indicated in Ref.~\cite{Huang:2011kf}, the phenomenological color screening confinement is an effective description of the hidden color channel coupling, so the hidden color channels of the pentaquark system in QDCSM are excluded.

The Goldstone-boson exchange interactions between light quarks appear because of the dynamical breaking of chiral symmetry. The following $\pi$, $K$, and $\eta$ exchange term work between the chiral quark-(anti)quark pair.
\begin{eqnarray*}
V_{\chi}(r_{ij}) & =&  v^{\pi}_{ij}(r_{ij})\sum_{a=1}^{3}\lambda_{i}^{a}\lambda_{j}^{a}+v^{K}_{ij}(r_{ij})\sum_{a=4}^{7}\lambda_{i}^{a}\lambda_{j}^{a}+v^{\eta}_{ij}(r_{ij})\\
&&\left[\left(\lambda _{i}^{8}\cdot
\lambda _{j}^{8}\right)\cos\theta_P-\left(\lambda _{i}^{0}\cdot
\lambda_{j}^{0}\right) \sin\theta_P\right], \label{sala-Vchi1}
\end{eqnarray*}
with
\begin{eqnarray*}
  v^{B}_{ij} &=&  {\frac{g_{ch}^{2}}{{4\pi}}}{\frac{m_{\chi}^{2}}{{\
12m_{i}m_{j}}}}{\frac{\Lambda _{\chi}^{2}}{{\Lambda _{\chi}^{2}-m_{\chi}^{2}}}}
m_{\chi}         \\
&&\left\{(\boldsymbol{\sigma}_{i}\cdot\boldsymbol{\sigma}_{j})
\left[ Y(m_{\chi}\,r_{ij})-{\frac{\Lambda_{\chi}^{3}}{m_{\chi}^{3}}}
Y(\Lambda _{\chi}\,r_{ij})\right] \right\},
B=\pi, K,  \eta,
\end{eqnarray*}

where $Y(x)=e^{-x}/x$ is the standard Yukawa function. The $\boldsymbol{\lambda^{a}}$ is the SU(3) flavor Gell-Mann matrix. The masses of the $\eta$, $K$ and $\pi$ meson are taken from the experimental value~\cite{ParticleDataGroup:2018ovx}. Finally, the chair coupling constant, $g_{ch}$, is determined from the $\pi NN$ coupling constant through
\begin{equation}
\frac{g_{ch}^{2}}{4\pi}=\left(\frac{3}{5}\right)^{2} \frac{g_{\pi NN}^{2}}{4\pi} {\frac{m_{u,d}^{2}}{m_{N}^{2}}},
\end{equation}
which assumes that flavor SU(3) is an exact symmetry, only broken by the different masses of the strange quark. Besides, with the Minuit program, we can determine a set of optimized parameters to fit ground state meson and baryon spectrum in QDCSM, which is shown in Table.~\ref{biaoge}. Then with the adjustable model parameters, the masses spectrum of meson and baryon can be obtained, which is listed in Table.~\ref{mass}.

\begin{table}[ht]
\caption{\label{biaoge}The values of the Model parameters. The masses of mesons take their experimental values.}
\renewcommand\arraystretch{1.25}
\begin{tabular}{p{2.5cm}<\centering p{2.5cm}<\centering p{2.5cm}<\centering }
 \toprule[1pt]
      & Parameter  &Value   \\
      \midrule[1pt]
Quark masses  &$m_u(MeV)$                      & 313 \\
              &$m_s(MeV)$                      & 573 \\
              &$m_c(MeV)$                      & 1788 \\
      \midrule[1pt]
confinement   &$b(fm)$                          &0.518\\
              &$a_{c}$(MeV $fm^{-2}$)           &58.03 \\
              &$V_{0_{qq}}(fm^2)$               &-1.2883\\
              &$V_{0_{q\bar{q}}}(fm^2)$         &-0.7432\\
     \midrule[1pt]
OGE           &$\alpha_{uu}$                     &0.5652\\
              &$\alpha_{us}$                     &0.5239\\
              &$\alpha_{uc}$                     &0.4673\\
              &$\alpha_{u\bar{s}}$               &1.4275\\
              &$\alpha_{s\bar{c}}$               &1.1901\\
    \midrule[1pt]
Goldstone boson     & $m_\pi(fm^{-1})$ & 0.7 \\
                    & $m_K (fm^{-1})$ & 2.51 \\
                    & $m_\eta(fm^{-1})$  & 2.77\\
                    & $\Lambda_{\pi}(fm^{-1})$ &4.2\\
                    & $\Lambda_{\eta/K }(fm^{-1})$    &5.2\\

\bottomrule[1pt]
\end{tabular}
\label{parameters}
\end{table}

\begin{table}[ht]
\caption{The Masses (in MeV) of the ground mesons. Experimental values are taken
from the Particle Data Group (PDG)~\cite{ParticleDataGroup:2018ovx}.}
\renewcommand\arraystretch{1.25}
\begin{tabular}{p{1cm}<\centering p{1cm}<\centering  p{1cm}<\centering p{1cm}<\centering p{1cm}<\centering p{1cm}<\centering p{1cm}<\centering p{1cm}<\centering }
\toprule[1pt]
\multicolumn{1}{c}{} &$N$  &$\Delta$ &$\Lambda$ &$\Sigma$ &$\Sigma^{\ast}$ &$\Xi$  &$\Xi^{\ast}$ \\
\midrule[1pt]
Expt                 &939  &1232     &1122      &1237     &1360             &1374  &1496      \\
Model                &939  &1232     &1116      &1189     &1385             &1318  &1533      \\
\midrule[1pt]
\multicolumn{1}{c}{} &$\Lambda_{c}$  &$\Sigma_{c}$ &$\Sigma_{c}^{\ast}$ &$K$ &$K^{\ast}$ &$D_{s}$ &$D_{s}^{\ast}$ \\
\midrule[1pt]
Expt                 &2286         &2464          &2489       &495          &815       &2018    &2064        \\
Model                &2286         &2455          &2520       &495          &895       &1968    &2112 \\
\bottomrule[1pt]
\end{tabular}
\label{mass}
\end{table}

In QDCSM, the quark delocalization is realized by specifying the single particle orbital
wave function of QDCSM as a linear combination of left and right Gaussian, the single
particle orbital wave functions used in the ordinary quark cluster model,
\begin{eqnarray}\label{wave0}
\psi_{r}(\boldsymbol{r},\boldsymbol{s}_{i},\epsilon)&=&\left(\phi_{R}(\boldsymbol{r},\boldsymbol{s}_{i})
  +\epsilon\phi_{L}(\boldsymbol{r},\boldsymbol{s}_{i})\right)/N(\epsilon), \nonumber \\
\psi_{l}(\boldsymbol{r},\boldsymbol{s}_{i},\epsilon)&=&\left(\phi_{L}(\boldsymbol{r},\boldsymbol{s}_{i})
  +\epsilon\phi_{R}(\boldsymbol{r},\boldsymbol{s}_{i})\right)/N(\epsilon), \nonumber \\
N(\epsilon)&=& \sqrt{1+\epsilon^2+2\epsilon e^{-s^2_{i}/{4b^2}}},\nonumber \\
\Phi_{\alpha}(\boldsymbol{s}_{i})&=&\left(\frac{1}{\pi b^2}\right)^{\frac{3}{4}}
e^{-\frac{1}{2b^2}\left(\boldsymbol{r_\alpha}-\frac{2}{5}s_{i}\right)^2},\nonumber \\
\Phi_{\beta}(-\boldsymbol{s}_{i})&=&\left(\frac{1}{\pi b^2}\right)^{\frac{3}{4}}
e^{-\frac{1}{2b^2}\left(\boldsymbol{r_\beta}+\frac{3}{5}s_{i}\right)^2},
\end{eqnarray}
The $\boldsymbol{s}_{i}$, $i=1,2,..., n$, are the generating coordinates, which are introduced to
expand the relative motion wave function~\cite{Wu:1998wu,Ping:1998si,Pang:2001xx}. The mixing parameter
$\epsilon(s_{i})$ is not an adjusted one but determined variationally by the dynamics of the
multi-quark system itself. This assumption allows the multi-quark system to choose its
favorable configuration in the interacting process. It has been used to explain the cross-over
the transition between the hadron phase and the quark-gluon plasma phase~\cite{Xu:2007oam}.

\section{The results and discussions\label{dis}}
In this work, we perform a systematical investigation of the low-lying charmed-strange pentaquark systems within the quark delocalization color screening model above. For the $nnnc\bar{s}$ (n: u, d) pentaquark system, we only consider the $s-$ wave channels with spin $S=\frac{1}{2}, \frac{3}{2}$ and $\frac{5}{2}$. All the channels involved are listed in Table~\ref{channe}. Our purpose of this work is to explore if there is any other pentaquark state and to see whether those pentaquark states can be explained as molecular pentaquarks.

\begin{table}[htb]
\begin{center}
\caption{\label{channels} The relevant channels for all possible states with different $J^P$ quantum numbers}
\renewcommand\arraystretch{1.5}
\begin{tabular}{p{1.0cm}<\centering p{0.8cm}<\centering p{0.8cm}<\centering p{0.8cm}<\centering p{0.8cm}<\centering| p{0.8cm}<\centering p{0.8cm}<\centering p{0.8cm}<\centering p{0.8cm}<\centering p{1.0cm}<\centering p{1.0cm}<\centering p{1.0cm}<\centering p{1.0cm}<\centering p{1.0cm}<\centering p{1.0cm}<\centering p{1.0cm}<\centering p{1.0cm}<\centering p{1.0cm}<\centering}
\toprule[1pt]
  &    \multicolumn{4}{c}{$I=\frac{1}{2}$}  & \multicolumn{4}{c}{$I=\frac{3}{2}$}\\
\midrule[1pt]
$S=\frac{1}{2}$ & $ND_{s}$ & $ND_{s}^{\ast}$ & $\Lambda_{c}K$ &$\Lambda_{c}K^{\ast}$ &$\Delta D_{s}^{\ast}$ &$\Sigma_{c}K$  &$\Sigma_{c}K^{\ast}$ &$\Sigma_{c}^{\ast}K^{\ast}$ \\
                &$\Sigma_{c}K$  &$\Sigma_{c}K^{\ast}$ &$\Sigma_{c}^{\ast}K^{\ast}$ &  &  &  \\
$S=\frac{3}{2}$ & $ND_{s}^{\ast}$ &$\Lambda_{c}K^{\ast}$  &$\Sigma_{c}K^{\ast}$  &$\Sigma_{c}^{\ast}K$ &$\Delta D_{s}$  &$\Delta D_{s}^{\ast}$ &$\Sigma_{c}K^{\ast}$ &$\Sigma_{c}^{\ast}K$   \\
                &$\Sigma_{c}^{\ast}K^{\ast}$ & &  &  &$\Sigma_{c}^{\ast}K^{\ast}$\\
$S=\frac{5}{2}$ &$\Sigma_{c}^{\ast}K^{\ast}$ & &  &  & $\Delta D_{s}^{\ast}$ &$\Sigma_{c}^{\ast}K^{\ast}$ \\
\bottomrule[1pt]
\end{tabular}
\end{center}
\label{channe}
\end{table}


\subsection{The effective potentials calculation}
\begin{figure}
\includegraphics[scale=0.55]{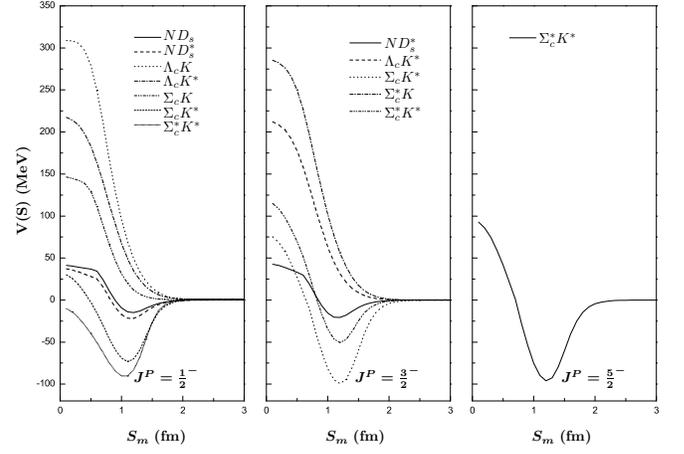}
 \caption{The effective potentials of different channels of the charmed-strange pentaquark systems with $I=\frac{1}{2}$ in QDCSM.}
\label{Veff-0.5}
\end{figure}

\begin{figure}
\includegraphics[scale=0.55]{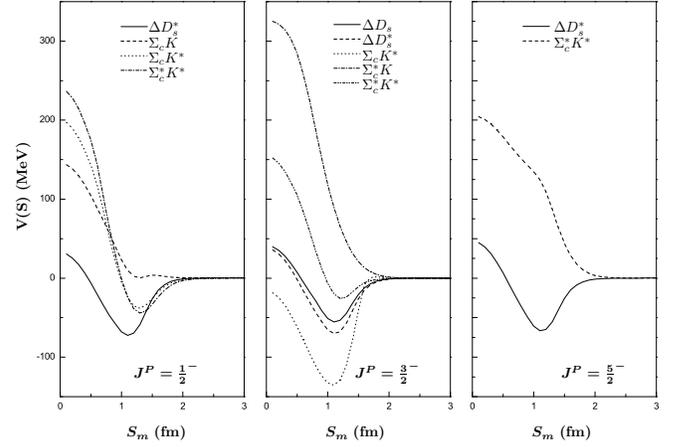}
 \caption{The effective potentials of different channels of the charmed-strange pentaquark systems with $I=\frac{3}{2}$ in QDCSM.}
\label{Veff-1.5}
\end{figure}

Because an attractive potential is necessary for forming a bound state or resonance state, so we first calculate the effective potentials between two hadrons, which are shown in Fig.~\ref{Veff-0.5} and Fig.~\ref{Veff-1.5}, respectively. The definition of the effective potential is $V(S)=E(S)-E(\infty)$ where $S$ represents the distance between two clusters, $E(S)$ stands for the energy of the system at the separation $S$ of two clusters, and $E(\infty)$ means the energy at a sufficient distance from $S$.

From Fig.~\ref{Veff-0.5}, some attractive potential channels can be found, which may be bound states or resonance states by the bound calculation and scattering calculation. For the $I(J^{P})=\frac{1}{2}(\frac{1}{2}^{-})$,  the $\Sigma_{c}^{\ast}K^{\ast}$, $\Sigma_{c}K^{\ast}$, $ND_{s}$ and $ND^{\ast}_{s}$ channels show attraction potential while $\Lambda_{c}K$, $\Lambda_{c}K^{\ast}$ and $\Sigma_{c}K$ channels have repulsive properties. In addition,  the $\Sigma_{c}^{\ast}K^{\ast}$ and $\Sigma_{c}K^{\ast}$ channels have deep attraction compared to the attraction potential of $ND_{s}$ and $ND^{\ast}_{s}$ channels, which indicates that the $\Sigma_{c}^{\ast}K^{\ast}$ and $\Sigma_{c}K^{\ast}$ are more likely to be bound states or resonant states. For the $I(J^{P})=\frac{1}{2}(\frac{3}{2}^{-})$, the potential of the $\Sigma_{c}K^{\ast}$, $\Sigma_{c}^{\ast}K^{\ast}$ and $ND_{s}^{\ast}$ channels show the attractive property while other two channels are repulsive. the attraction of the $\Sigma_{c}$ and $K^{\ast}$ is much larger than that of $\Sigma_{c}^{\ast}K^{\ast}$ and $ND_{s}^{\ast}$, which implies that it is possible for $\Sigma_{c}K^{\ast}$ to form a bound or resonance state. For the $I(J^{P})=\frac{1}{2}(\frac{5}{2}^{-})$, the only channel $\Sigma_{c}^{\ast}K^{\ast}$ has a strong attraction, so this channel may be also a bound state or resonance state.

For the $I=\frac{3}{2}$ system, from Fig.~\ref{Veff-1.5}, the results of the effective potential are similar to the results of the $I=\frac{1}{2}$ system. For the $J^{P}=\frac{1}{2}^{-}$, one can see that the potentials are all attractive for the channels $\Delta D_{s}^{\ast}$, $\Sigma_{c}K^{\ast}$ and $\Sigma_{c}^{\ast}K^{\ast}$. For $\Sigma_{c}K$ channel, the potential of which is repulsive, so no bound state or resonance states can be formed in this channel. However, the bound states or resonance states are possible for other channels due to the attractive nature of the interaction between the two clusters. From Fig.~\ref{Veff-1.5}, the attraction of the $\Delta$ and $D_{s}^{\ast}$ is the largest one, followed by $\Sigma_{c}^{\ast}K^{\ast}$ channel and $\Sigma_{c}K^{\ast}$ channel, the attraction of these two effective potentials is almost the same. For the $J^{P}=\frac{3}{2}^{-}$, the interaction of all the channels except $\Sigma_{c}^{\ast}K$ channel is attractive.  The $\Sigma_{c}K^{\ast}$ has a very strong attraction with maximum attraction energy of $-148$ MeV, followed by the $\Delta D_{s}^{\ast}$ and $\Delta D_{s}$ with the largest attraction energy of $-55$ MeV and $-51$ MeV. However, the attraction of the $\Sigma_{c}^{\ast}K^{\ast}$ is the smallest one with the largest attraction energy of $-6$ MeV. For the $J^{P}=\frac{5}{2}^{-}$, there are two channel $\Delta D_{s}^{\ast}$ and $\Sigma_{c}^{\ast}K^{\ast}$, the potential of the $\Delta D_{s}^{\ast}$ is attractive while the potential of  $\Sigma_{c}^{\ast}K^{\ast}$ is repulsive.

\subsection{The bound-state calculation}
In this section, the bound state calculation is mainly performed for the $nnnc\bar{s}$ pentaquark system based on the resonating group method (RGM)~\cite{Kamimura:1981oxj,Kamimura:1977oxj}. The purpose is to confirm whether the states with attractive interaction can form bound states. The results are listed in Table~\ref{bound-1} and Table~\ref{bound-2}. The first two columns represent the quantum number of the system and the corresponding state of every channel. The $E_{sc}$ denotes the eigenenergy obtained for every channel. The $E_{th}$ and $E_{exp}$ stand for the sum of  the theoretical and experimental thresholds for baryons and mesons. $E_{B}$ is the binding energy, which can be obtained by the difference of $E_{sc}$ and $E_{th}$ $(E_{B}=E_{sc}-E_{th})$. Here, it is important to note that the bound state is present when $E_{B}<0$. The corrected energy $E_{sc}^{\prime}$ can be available through the sum of $E_{exp}$ and $E_{B}$ while $E_{cc}$ is the lowest energy of the system by channel-coupling calculation. By treating the data in this way we hope partially reduce the errors introduced by the model parameters.

For the system with $I(J^{P})=\frac{1}{2}(\frac{1}{2}^{-})$, the single channel calculation shows that only two channels $\Sigma_{c}K^{\ast}$ and $\Sigma_{c}^{\ast}K^{\ast}$ with the binding energy -7 MeV and -20 MeV are bound states, their eigenenergies are 3340 MeV and 3392 MeV, respectively. This result is consistent with the behavior of the effective potential. From Fig.~\ref{Veff-0.5}, $\Sigma_{c}K^{\ast}$ and $\Sigma_{c}^{\ast}K^{\ast}$ have strong attractive interaction to form bound states, so it is reasonable to obtain two bound states.
Also, comparing to the results of Ref.~\cite{Chen:2022svh} which concluded that the single channel $\Sigma_{c}K^{\ast}$ with $I(J^{P})=\frac{1}{2}(\frac{1}{2}^{-})$ can be good charmed-strange molecular state, its conclusion supports our result.
However, for the $ND_{s}$ and $ND_{s}^{\ast}$ states, the attraction for the $ND_{s}$ and $ND_{s}^{\ast}$ channels are too weak to form bound states. The other channels' energies are above the threshold of the corresponding physical channels due to the nature of repulsion. After coupling all channels, the obtained energy $2784$ MeV is still above the threshold of the lowest physical channel $\Lambda_{c}K$, so there is no bound state.
Moreover, the $\Sigma_{c}K^{\ast}$ and $\Sigma_{c}^{\ast}K^{\ast}$
states can decay to the corresponding open channels by coupling to open channels and it is possible for two bound states to become resonance states. To confirm the possibility that these two channels can form resonance states, the study of the scattering process of the open channels is needed, which is discussed in the next section.

For the system with $I(J^{P})=\frac{1}{2}(\frac{3}{2}^{-})$, there are five channels: $ND_{s}^{\ast}$, $\Lambda_{c}K^{\ast}$, $\Sigma_{c}K^{\ast}$, $\Sigma_{c}^{\ast}K$ and $\Sigma^{\ast}_{c}K^{\ast}$. The single-channel calculation indicates the $\Sigma_{c}K^{\ast}$ channel is bound with the binding energy of -23 MeV due to a deep attractive interaction between $\Sigma_{c}$ and $K^{\ast}$. Nevertheless $ND_{s}^{\ast}$ and $\Sigma^{\ast}_{c}K^{\ast}$ can not form bound states because their attractive interaction is too small. The lowest energy of this system is $3016$ MeV after the channel-coupling calculation, higher than the threshold of the lowest channel $\Sigma_{c}^{\ast}K$ as shown in Table~\ref{bound-1}. However, we still need to calculate the scattering process of the open channels to check if the $\Sigma_{c}K^{\ast}$ is resonance or not, these results will be shown in the next section.

For the system with $I(J^{P})=\frac{1}{2}(\frac{5}{2}^{-})$, it includes only channel $\Sigma^{\ast}_{c}K^{\ast}$. From Table.~\ref{bound-1}, the mass of $\Sigma^{\ast}_{c}K^{\ast}$ is 3392 MeV with the binding energy of -20 MeV, which indicates the attractive interaction between $\Sigma^{\ast}_{c}$ and $K^{\ast}$ is very strong. Besides, from Fig.~\ref{Veff-0.5}, the conclusions are reasonable and consistent.

For the system with $I(J^{P})=\frac{3}{2}(\frac{1}{2}^{-})$, the strong attractive interaction between $\Delta$ and $D_{s}^{\ast}$ leads to the obtained energy below the threshold of the two particles while the other remaining channels are above the threshold in the single-channel calculations as shown in Table~\ref{bound-2}. When considering the channel coupling calculation, the lowest energy of this system is pushed to 2952 MeV,  2 MeV higher than the threshold of the lowest channel $\Sigma_{c}K$, which means that the system with $I(J^{P})=\frac{3}{2}(\frac{1}{2}^{-})$ is unbound.

For the system with $I(J^{P})=\frac{3}{2}(\frac{3}{2}^{-})$, there are five channels as shown in Table~\ref{bound-2}. The single-channel calculation demonstrates that three states $\Delta D_{s}$, $\Delta D_{s}^{\ast}$ and $\Sigma_{c} K^{\ast}$ are bound and their binding energy are -1 MeV, -7 MeV, and -55 MeV, respectively. This result is consistent with the behavior of the effective potential. Besides, three bound states can decay to some open channel, and we can then determine the nature of these bound states in the scattering of these open channels, which are present in the next section. Moreover, the lowest energy of this system is 3019 MeV, which is higher than the threshold of the lowest physical channel $\Sigma_{c}^{\ast}K$ by the channel-coupling calculation. So the system  with $I(J^{P})=\frac{3}{2}(\frac{3}{2}^{-})$ is unbound. The results are similar to the results of systems with quantum numbers of $I(J^{P})=\frac{1}{2}(\frac{1}{2}^{-})$ and $\frac{1}{2}(\frac{3}{2}^{-})$.

For the system with $I(J^{P})=\frac{3}{2}(\frac{5}{2}^{-})$, in the single-channel calculation $\Delta D_{s}^{\ast}$ is bound state with a binding energy of -5 MeV, and the energy of $\Sigma^{\ast}_{c}K^{\ast}$ is 3417 MeV, which is 5 MeV higher than the corresponding threshold $3412$ MeV, so $\Sigma^{\ast}_{c}K^{\ast}$ is unbound. Besides, there is a bound state with the binding energy of -37 MeV by the channel-coupling calculation, which shows that the channel-coupling calculation is important for the existence of bound states.

\begin{figure}
\includegraphics[scale=0.5]{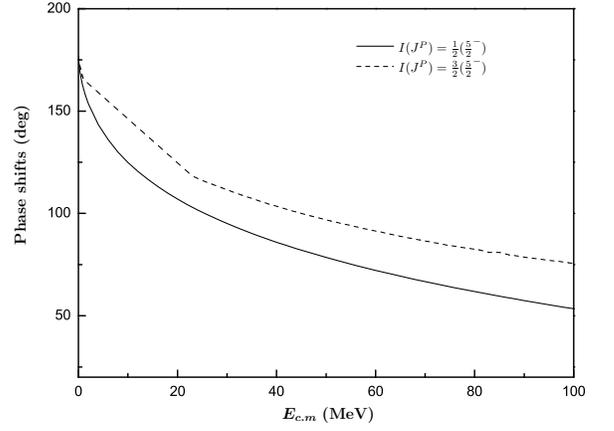}
 \caption{The phase shifts of the $I(J^{P})=\frac{1}{2}(\frac{5}{2}^{-})$ $\Sigma_{c}^{\ast}K^{\ast}$ and $I(J^{P})=\frac{3}{2}(\frac{5}{2}^{-})$ $\Delta D_{s}^{\ast}$ with channel coupling .}
\label{low-phase}
\end{figure}

According to the bound calculation, we obtain two bound states in $I(J^{P})=\frac{1}{2}(\frac{5}{2}^{-})$ and $\frac{3}{2}(\frac{5}{2}^{-})$ system. To further check the possibility of those bound states, the low-energy scattering phase shifts of which are investigated by the variational method.  The details of this method can be found in Appendix~\ref{app}. The scattering length $a_{0}$, the effective range $r_{0}$, and the binding energy $E_{B}^{\prime}$ are calculated and these results are all shown in Table.~\ref{bound-3}, which confirm the existence of bound states. Moreover, from the Fig.~\ref{low-phase}, it is obvious that the low-energy phase shifts of the $I(J^{P})=\frac{1}{2}(\frac{5}{2}^{-})$ $\Sigma_{c}^{\ast}K^{\ast}$ and $I(J^{P})=\frac{3}{2}(\frac{5}{2}^{-})$ $\Delta D_{s}^{\ast}$ with channel coupling can up to 180 degrees at $E_{c.m}\sim0$ and decreases rapidly as $E_{c.m}$ keeps increasing. This behavior of Fig.~\ref{low-phase} also indicates the existence of bound states. 

\begin{table}[htb]
\begin{center}
\renewcommand{\arraystretch}{1.5}
\caption{\label{bound-1} The binding energies and the masses of every single channel and those of channel coupling for the pentaquarks with $I=\frac{1}{2}$. The values are provided in units of MeV. }
\begin{tabular}{p{1.cm}<\centering p{1.cm}<\centering p{1.2cm}<\centering p{1.2cm}<\centering p{1.2cm}<\centering p{1.2cm}<\centering p{1.2cm}<\centering p{1.2cm}<\centering  }
\toprule[1pt]
$I(J^{P})$  & Channel & $E_{sc}$ &$E_{th}$ &$E_{B}$ &$E_{exp}$ &$E_{sc}^{\prime}$ \\
 \midrule[1pt]
\multirow{8}{*}{$\frac{1}{2}(\frac{1}{2}^{-})$} &$ND_{s}$   &2954 &2957 &$+3$ &2907 &2910 \\
                               &$ND_{s}^{\ast}$             &3006 &3003 &$+3$ &3051 &3054\\
                               &$\Lambda_{c}K$              &2785 &2781 &$+4$ &2781 &2785\\
                               &$\Lambda_{c}K^{\ast}$       &3104 &3100 &$+4$ &3178 &3182\\
                               &$\Sigma_{c}K$               &2963 &2959 &$+4$ &2950 &2954\\
                               &$\Sigma_{c}K^{\ast}$        &3277 &3278 &$-7$ &3347 &3340\\
                               &$\Sigma_{c}^{\ast}K^{\ast}$ &3283 &3303 &$-20$&3412 &3392\\
                               &$E_{cc}$                    &2784 &     &$+3$ &     &2784\\
\midrule[1pt]
\multirow{6}{*}{$\frac{1}{2}(\frac{3}{2}^{-})$} &$ND_{s}^{\ast}$             &3006 &3003 &$+3$ &3051 &3054 \\
                                                &$\Lambda_{c}K^{\ast}$       &3104 &3100 &$+4$ &3178 &3182\\
                                                &$\Sigma_{c}K^{\ast}$        &3255 &3278 &$-23$&3347 &3324\\
                                                &$\Sigma_{c}^{\ast}K$        &2989 &2984 &$+5$ &3015 &3020\\
                                                &$\Sigma_{c}^{\ast}K^{\ast}$ &3305 &3303 &$+2$ &3412 &3414\\
                                                &$E_{cc}$                    &2985 &     &$+1$ &     &3016\\
                                                \midrule[1pt]
\multirow{2}{*}{$\frac{1}{2}(\frac{5}{2}^{-})$} &$\Sigma_{c}^{\ast}K^{\ast}$ &3283 &3303 &$-20$ &3412 &3392\\
                                                &$E_{cc}$                    &3283 &     &$-20$ &     &3392\\
\bottomrule[1pt]
\end{tabular}
\end{center}
\end{table}

\begin{table}[htb]
\begin{center}
\renewcommand{\arraystretch}{1.5}
\caption{\label{bound-2} The binding energies and the masses of every single channel and those of channel coupling for the pentaquarks with $I=\frac{3}{2}$. The values are provided in units of MeV. }
\begin{tabular}{p{1.cm}<\centering p{1.cm}<\centering p{1.2cm}<\centering p{1.2cm}<\centering p{1.2cm}<\centering p{1.2cm}<\centering p{1.2cm}<\centering p{1.2cm}<\centering  }
\toprule[1pt]
$I(J^{P})$  & Channel & $E_{sc}$ &$E_{th}$ &$E_{B}$ &$E_{exp}$ &$E_{sc}^{\prime}$ \\
 \midrule[1pt]
\multirow{5}{*}{$\frac{3}{2}(\frac{1}{2}^{-})$} &$\Delta D_{s}^{\ast}$ &3288  &3296 &$-8$ &3344 &3336\\
                               &$\Sigma_{c}K$               &2963 &2959 &$+4$ &2950 &2954\\
                               &$\Sigma_{c}K^{\ast}$        &3281 &3278 &$+3$ &3347 &3350\\
                               &$\Sigma_{c}^{\ast}K^{\ast}$ &3306 &3303 &$+3$ &3412 &3415\\
                               &$E_{cc}$                    &2961 &     &$+2$ &     &2952\\
\midrule[1pt]
\multirow{6}{*}{$\frac{3}{2}(\frac{3}{2}^{-})$} &$\Delta D_{s}$              &3249 &3250 &$-1$ &3200 &3199 \\
                                                &$\Delta D_{s}^{\ast}$       &3289 &3296 &$-7$ &3344 &3337 \\
                                                &$\Sigma_{c}K^{\ast}$        &3223 &3278 &$-55$&3347 &3292\\
                                                &$\Sigma_{c}^{\ast}K$        &2989 &2984 &$+5$ &3015 &3020\\
                                                &$\Sigma_{c}^{\ast}K^{\ast}$ &3307 &3303 &$+4$ &3412 &3416\\
                                                &$E_{cc}$                    &2991 &     &$+4$ &     &3019\\
                                                \midrule[1pt]
\multirow{3}{*}{$\frac{3}{2}(\frac{5}{2}^{-})$} &$\Delta D_{s}^{\ast}$       &3291 &3296 &$-5$ &3344 &3339 \\
                                                &$\Sigma_{c}^{\ast}K^{\ast}$ &3308 &3303 &$+5$ &3412 &3417\\
                                                &$E_{cc}$                    &3259 &     &$-37$ &    &3307\\
\bottomrule[1pt]
\end{tabular}
\end{center}
\end{table}

\begin{table}[htb]
\begin{center}
\renewcommand{\arraystretch}{1.5}
\caption{\label{bound-3} The binding energy $E_{B}^{\prime}$ (MeV), the scattering length $a_{0}$(fm) and the effective range $r_{0}$(fm). }
\begin{tabular}{p{1.5cm}<\centering p{1.5cm}<\centering p{1.5cm}<\centering p{1.5cm}<\centering p{1.5cm}<\centering   }
\toprule[1pt]
$I(J^{P})$  & Channel & $a_{0}$ &$r_{0}$ &$E_{B}^{\prime}$ \\
\midrule[1pt]
$\frac{1}{2}(\frac{5}{2}^{-})$ & $\Sigma_{c}^{\ast}K^{\ast}$ &2.0625 &0.97682 & -23.8 \\
$\frac{3}{2}(\frac{5}{2}^{-})$ & $\Delta D_{s}^{\ast}$       &2.0032 &0.99889 & -34.7\\
\bottomrule[1pt]
\end{tabular}
\end{center}
\end{table}

\subsection{The resonance state calculation}
\begin{table}[htb]
\begin{center}
\renewcommand{\arraystretch}{1.5}
\caption{\label{R} The masses and decay widths of resonance states with the difference scattering process. $R^{\prime}$ stands for the modified resonance mass.  $\Gamma$ is the partial decay width of the resonance state decaying to an open channel. $\Gamma_{Total}$ is the total decay width of the resonance state.. }
\begin{tabular}{p{2.8cm}<\centering p{1.cm}<\centering p{1.cm}<\centering p{0.3cm}<\centering p{1.cm}<\centering p{1.cm}<\centering p{1.cm}<\centering p{1.2cm}<\centering  }
\toprule[1pt]
\multicolumn{1}{c}{} & \multicolumn{5}{c}{three-channel coupling} \\
\cline{2-6}
\multicolumn{1}{c}{}  &\multicolumn{2}{c}{$I(J^{P})=\frac{1}{2}(\frac{1}{2}^{-})$} & &\multicolumn{2}{c}{$I(J^{P})=\frac{1}{2}(\frac{1}{2}^{-})$} \\
 \cline{2-3}\cline{5-6}
 Open channels  &\multicolumn{2}{c}{$\Sigma_{c}K^{\ast}$} &  &\multicolumn{2}{c}{$\Sigma_{c}^{\ast}K^{\ast}$}\\
 \midrule[1pt]
 \multicolumn{1}{c}{}                     &$R^{\prime}$       &$\Gamma$            &            &$R^{\prime}$    & $\Gamma$\\
  \cline{2-3}\cline{5-6}
     $ND_{s}$                             &$\cdots$  &$\cdots$            &            &$\cdots$ &$\cdots$ \\
     $ND_{s}^{\ast}$                      &3344      &5.1                 &            &$\cdots$ &$\cdots$ \\
     $\Lambda_{c}K$                       &$\cdots$  &$\cdots$            &            &$\cdots$ &$\cdots$ \\
     $\Lambda_{c}K^{\ast}$                &3346      &1.4                 &            &$\cdots$ &$\cdots$ \\
     $\Sigma_{c}K$                        &3342      &19                  &            &$\cdots$ &$\cdots$ \\
     $\Gamma_{Total}$                       &          &25.5                &            &         &         \\
  \midrule[1pt]
  \multicolumn{1}{c}{} & \multicolumn{5}{c}{two-channel coupling} \\
\cline{2-6}
\multicolumn{1}{c}{}  &\multicolumn{2}{c}{$I(J^{P})=\frac{1}{2}(\frac{3}{2}^{-})$} & &\multicolumn{2}{c}{$I(J^{P})=\frac{3}{2}(\frac{1}{2}^{-})$} \\
 \cline{2-3}\cline{5-6}
 Open channels  &\multicolumn{2}{c}{$\Sigma_{c}K^{\ast}$} &  &\multicolumn{2}{c}{$\Delta D_{s}^{\ast}$}\\
 \midrule[1pt]
 \multicolumn{1}{c}{}                     &$R^{\prime}$ & $\Gamma$                &            &$R^{\prime}$    & $\Gamma$\\
  \cline{2-3}\cline{5-6}
     $ND_{s}^{\ast}$                       &$\cdots$  &$\cdots$          &            &$\cdots$ &$\cdots$ \\
     $\Lambda_{c}K^{\ast}$                 &$\cdots$  &$\cdots$          &            &$\cdots$ &$\cdots$ \\
     $\Sigma_{c}       K$                  &$\cdots$  &$\cdots$          &            &3343     &0.01 \\
     $\Sigma_{c}^{\ast}K$                  &3333      &3.3               &            &$\cdots$ &$\cdots$ \\
     $\Gamma_{Total}$                      &          &3.3               &            &         &0.01         \\
\bottomrule[1pt]
\end{tabular}
\end{center}
\end{table}

\begin{figure*}
\includegraphics[scale=2.0]{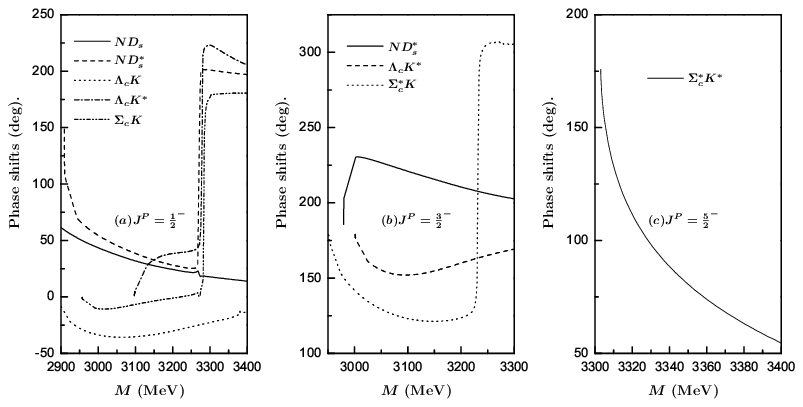}
 \caption{The phase shifts of the open channels with $I=\frac{1}{2}$ in QDCSM. M(MeV) is the sum of the corresponding theoretical threshold of the open channel and the incident energy. }
\label{R-0.5}
\end{figure*}

\begin{figure*}
\includegraphics[scale=2.0]{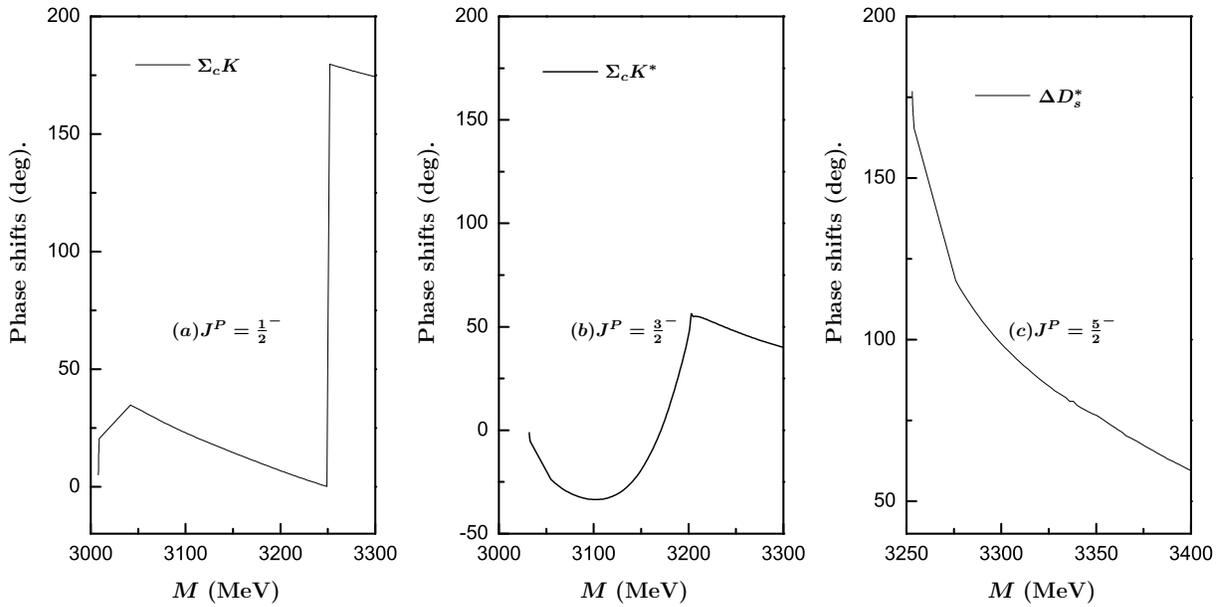}
 \caption{The phase shifts of the open channels with $I=\frac{3}{2}$ in QDCSM. M(MeV) is the sum of the corresponding theoretical threshold of the open channel and the incident energy. }
\label{R-1.5}
\end{figure*}

According to the results above, some bound states can be obtained in the single-channel calculation due to the attractive interaction between the two hadrons. These states can decay to the corresponding open channels by coupling to open channels and may become resonance states, but it is not excluded that these states become scattered states under the coupling effect of open channels and closed channels. So to determine whether resonance states would exist, we studied the scattering phase shifts of all possible open channels in QDCSM, and the scattering phase shifts behavior of open channels is shown in Fig.~\ref{R-0.5} and Fig.~\ref{R-1.5}. In addition, the mass and the decay width of the possible resonance states are also calculated and listed in Table.~\ref{R}. The current calculation applies only to the decay of $s-$wave open channels due to the extremely small and almost negligible decay of the higher fractional waves.

For the $I(J^{P})=\frac{1}{2}(\frac{1}{2}^{-})$, the information obtained from the bound energy calculation demonstrates that two bound states, $\Sigma_{c}K^{\ast}$ and $\Sigma_{c}^{\ast}K^{\ast}$, are available in the single channel calculation. According to the Table~\ref{bound-1}, the possible decay channels of these bound states may be $ND_{s}$, $ND_{s}^{\ast}$, $\Lambda_{c}K$, $\Lambda_{c}K^{\ast}$ and $\Sigma_{c}K$. So we analyze the scattering phase shifts of three-channel coupling with two bound states and corresponding open channels. From Fig.~\ref{R-0.5}(a), there is no resonance state in the scattering phase shifts of $ND_{s}$ and $\Lambda_{c}K$ but the phase shifts of the $ND_{s}^{\ast}$, $\Lambda_{c}K^{\ast}$ and $\Sigma_{c}K$ channels shows one resonance state, which means that a bound state $\Sigma_{c}K^{\ast}$ appears as a resonance state by three-channel coupling. On the contrary, the $\Sigma_{c}^{\ast}K^{\ast}$ vanishes due to the channel coupling effect pushing it above the threshold. The resonance mass and decay width can be obtained from the shape of the resonance. Here, in order to minimize the theoretical errors and to compare calculated results to future experimental data, we shift the resonance mass to $R^{\prime}=M-E_{th}+E_{exp}$. Taking the resonance state $\Sigma_{c}K^{\ast}$ in scattering channel $\Sigma_{c}K$ as an example, the resonance mass $M=3273$ MeV, from Table.~\ref{bound-1}, $E_{th}=3278$ MeV and $E_{exp}=3347$ MeV, then finally the modified resonance mass $R^{\prime}=3273-3278+3347=3342$ MeV. The results of the resonance mass and the decay width are listed in Table.~\ref{R}. The resonance mass range of the $\Sigma_{c}K^{*}$ is $3342\sim3346$ MeV and the total decay width is about $25.5$ MeV. From the Table.~\ref{R}, it can be found that the mass shifts of the resonance mass are small, which indicates that the scattering channel and bound state channel coupling effect is not strong due to the large mass difference between the scattering channel and bound state channel.

For the $I(J^{P})=\frac{1}{2}(\frac{3}{2}^{-})$, a bound state $\Sigma_{c}K^{\ast}$ can decay to the open channel $ND_{s}^{\ast}$, $\Lambda_{c}K^{\ast}$ and $\Sigma_{c}^{\ast} K$, respectively. So the scattering phase shift two-channel coupling with a bound state channel and the open channel is calculated and the results are shown in Fig.~\ref{R-0.5}(b). The resonance state $\Sigma_{c}K^{\ast}$ can be found in the scattering phase shift of $\Sigma_{c}^{\ast} K$ but not in the remaining scattering phase shift. From Table~\ref{R}, the resonance mass and decay widths are 3333 MeV and 3.1 MeV, respectively.  However, for the $I(J^{P})=\frac{1}{2}(\frac{5}{2}^{-})$, from Fig.~\ref{R-0.5}(b), the only one channel $\Sigma_{c}^{\ast}K^{\ast}$ is close to 180 as $M\sim 0$, which means that this state is a bound state and not a resonant state when we only only consider the scattering process under s-wave.

For the $I=\frac{3}{2}$ system, the behavior of the scattering phase shift of the open channel is presented in Fig.~\ref{R-1.5}. First, we analyze the scattering phase shift of $J^{P}=\frac{1}{2}^{-}$, there is only a bound state $ \Delta D_{s}^{\ast}$ in the single channel calculation with the binding energy of $-8$ MeV. This bound state can decay to the open channel $\Sigma_{c}K$, form the Fig.~\ref{R-1.5}(a), the resonance state $\Delta D_{s}^{\ast}$ can be obtained in the scattering phase shift of $\Sigma_{c}K$. From the Tabel~\ref{R}, the resonance mass of $\Delta D_{s}^{\ast}$ in the $I(J^{P})=\frac{3}{2}(\frac{1}{2}^{-})$ is about 3343 MeV and decay width is approximately 0.01 MeV. For the $J^{P}=\frac{3}{2}^{-}$, three bound states $\Delta D_{s}$, $\Delta D_{s}^{\ast}$ and $\Sigma_{c} K^{\ast}$ can only decay to the open channel $\Sigma_{c}^{\ast} K$. So in the present work four-channel coupling is considered. From Fig.~\ref{R-1.5}(b), no resonance states appear in the scattering phase shift of $\Sigma_{c}^{\ast} K$, which reveals that the bound states $\Delta D_{s}$, $\Delta D_{s}^{\ast}$ and $\Sigma_{c} K^{\ast}$ become scattering states through the four-channel coupling. For the $J^{P}=\frac{5}{2}^{-}$, although the $\Delta D_{s}^{\ast}$ is a bound state in the single channel calculation,  this channel cannot decay to  the $\Sigma_{c}^{\ast} K^{\ast}$ because the experiment of $\Sigma_{c}^{\ast} K^{\ast}$ is higher than that of $\Delta D_{s}^{\ast}$. The phase shifts of $\Delta D_{s}^{\ast}$ is presented in Fig.~\ref{R-1.5}(c), the behavior of which is similar to that of the $\Sigma_{c}^{\ast}K^{\ast}$ with $I(J^{P})=\frac{1}{2}(\frac{5}{2}^{-})$.

\section{Summary\label{sum}}
Recently, the LHCb Collaboration reported the observation of the $T_{c\bar{s}0}^{a}(2900)^{++}$ and $T_{c\bar{s}0}^{a}(2900)^{0}$ in the $D_{s}^{+}\pi^{-(+)}$ invariant mass spectrums. Inspired by these newly discovered exotic states, it is possible that similar charmed-strange pentaquark states with quark composed of $nnnc\bar{s}$ exist in the real physical world. In this work the charmed-strange pentaquark system is investigated by using the resonance group method in the framework of QDCSM that satisfactorily describes the $P_{\psi}^{N}$~\cite{}. Herein, the effective potential calculations are performed to explore the interactions of individual channels in different quantum number systems; the single-channel calculation and channel coupling calculation are carried out to find the bound states; besides, the scattering phase shifts calculation is made to discover possible resonance states.

the numerical results show that for the charmed-strange pentaquark system, first there are two bound states $\Sigma_{c}^{\ast}K^{\ast}$ with $I(J^{P})=\frac{1}{2}(\frac{5}{2}^{-})$ and $\Delta D_{s}^{\ast}$ with $I(J^{P})=\frac{3}{2}(\frac{5}{2}^{-})$ through channel coupling calculation. Apart from the bound states, some resonance states can be obtained in QDCSM. The resonance state $\Sigma_{c}K^{\ast}$ with $I(J^{P})=\frac{1}{2}(\frac{1}{2}^{-})$ can be found in the scattering phase shifts of $N D_{s}^{\ast}$, $\Lambda_{c}K^{\ast}$ and $\Sigma_{c}K$, and the resonance mass $(R^{\prime})$ and the total decay width $(\Gamma_{Total})$ are about $3342\sim3346$ MeV and 25.5 MeV. The $\Sigma_{c}K^{\ast}$ with $I(J^{P})=\frac{1}{2}(\frac{3}{2}^{-})$ is seen in the $\Sigma_{c}^{\ast}K$ channel. The resonance mass is about 3333 MeV and the decay width is about 3.3 MeV. In addition, by coupling the $\Sigma_{c}K$ channel the resonance state $\Delta D^{\ast}$ with $I(J^{P})=\frac{3}{2}(\frac{1}{2}^{-})$ is also visible in the scattering phase shift of the $\Sigma_{c}K$ channel, so the mass and the total decay width of the resonance state $\Delta D^{\ast}$ are 3343 MeV and 0.01 MeV, respectively.

\acknowledgments{This work is supported partly by the National Natural Science Foundation of China under
Contract Nos. 12175037, 11775050, 11775118, 11535005, and 11865019, and is also supported by the Fundamental Research Funds for the Central Universities No. 2242022R20040; and the China Postdoctoral Science Foundation No.2021M690626 and No.1107020201.Besides, Jiangsu Provincial Natural Science Foundation Project,No.BK20221166 and National Youth Fund:No.12205125 are also supported this work.}

\appendix
\section{resonating group method for bound-state and scattering problems \label{app}}
In the present work, We perform bound state calculations as well as scattering calculations for the $N\bar{D_{s}}$ system using the RGM. The core of this approach is how to deal with the two-body problem. In this method, when dealing with the two-cluster system, one can only consider the relative motion between the clusters, while the two clusters are frozen inside. So the wave function of the baryon-meson system is
\begin{eqnarray}\label{wave1}
    \psi &=& \sum_{L}\mathcal{A}\left[\left[\hat{\phi_{A}}\left(\boldsymbol{\rho_{A}},\boldsymbol{\lambda_{A}}\right)\hat{\phi_{B}}\left(\boldsymbol{\rho_{B}}\right)\right]^{[\sigma]IS}\otimes \chi_{L}(\textbf{R}_{AB})\right]^{J},
\end{eqnarray}
where the symbol $\mathcal{A}$ is the antisymmetry operator, which can be defined as
\begin{eqnarray}\label{wave2}
    \mathcal{A}&=& 1-P_{14}-P_{24}-P_{34},
\end{eqnarray}
where 1, 2, and 3 stand for the quarks in the baryon cluster, and 4 stands for the quark in the meson cluster. $\hat{\phi_{A}}$ and $\hat{\phi_{B}}$ are the internal cluster wave functions of the baryon A and meson B:
\begin{eqnarray}\label{wave3}
 \hat{\phi_{A}} &=& \left(\frac{2}{3\pi b^{2}}\right)^{3/4} \left(\frac{1}{2\pi b^{2}}\right)^{3/4}  e^{-\left(\frac{\boldsymbol{\rho_{A}}^2}{4b^2}+\frac{\boldsymbol{\lambda_{A}}^2}{3b^2}\right)} \eta_{I_{A}}S_{A}\chi_{A}^{c} ,\\
 \hat{\phi_{B}} &=& \left(\frac{1}{2\pi b^{2}}\right)^{3/4} e^{-\frac{\boldsymbol{\rho_{B}}}{4b^2}} \eta_{I_{B}}S_{B}\chi_{B}^{c} ,
\end{eqnarray}
where $\eta_{I}$, $S$, and $\chi$ represent the flavor, spin, and internal color terms of the cluster wave functions, respectively. $\rho_{A}$ and $\lambda_{A}$ are the internal coordinates for the baryon cluster A and $\rho_{B}$ is the internal coordinate for the meson cluster B. The Jacobi coordinates are defined as follows:
\begin{eqnarray}\label{wave4}
\boldsymbol{\rho_{A}}&=&\boldsymbol{r_{1}-r_{2}}, \ \ \ \boldsymbol{\rho_{B}}=\boldsymbol{r_{4}-r_{5}},\nonumber \\
\boldsymbol{\lambda_{A}}&=& \boldsymbol{r_{3}}-\frac{1}{2}(\boldsymbol{r}_{1}+\boldsymbol{r}_{2}),\nonumber \\
\boldsymbol{R_{A}}&=& \frac{1}{3}(\boldsymbol{r_{1}}+\boldsymbol{r_{2}}+\boldsymbol{r_{3}}), \ \ \ \boldsymbol{R_{B}}=\frac{1}{2}(\boldsymbol{r_{4}}+r_{5}),\nonumber \\
\boldsymbol{R_{AB}}&=&\boldsymbol{R_{A}}-\boldsymbol{R_{B}},\ \ \ \boldsymbol{R_{G}}=\frac{3}{5}\boldsymbol{R_{A}}+\frac{2}{5}\boldsymbol{R_{B}}.
\end{eqnarray}
From the variational principle, after variation with respect to the relative motion wave function $\chi\boldsymbol(R)=\sum_{L}\chi_{L}\boldsymbol(R)$, one obtains the RGM equation
\begin{eqnarray}\label{wave5}
  \int H\left(\boldsymbol{R, R^{\prime}}\right)\chi\left(\boldsymbol{R^{\prime}}\right)d\boldsymbol\left(R^{\prime}\right)=E\nonumber \\  \int N\left(\boldsymbol{R, R^{\prime}}\right) \chi\left(\boldsymbol{R^{\prime}}\right)d\boldsymbol\left(R^{\prime}\right),
\end{eqnarray}
where $H(\boldsymbol{R, R^{\prime}})$ and $N(\boldsymbol{R, R^{\prime}})$ are Hamiltonian and norm kernels, respectively. The eigenenergy $E$ and the wave functions are obtained by solving the RGM equation. In the present estimation, we introduce generator coordinates $S_{m}$ to expand the $L$th relative motion wave function $\chi_{L}(\boldsymbol{R})$:
 \begin{eqnarray}\label{wave6}
\chi_{L}\boldsymbol{(R)}&=&\frac{1}{\sqrt{4\pi}}\left(\frac{6}{5\pi b^2}\right)^{3/4}\sum_{m=1}^{n}C_{m}\nonumber\\
&&\int exp\left[-\frac{3}{5b^2}\left(\boldsymbol{R}-\boldsymbol{S}_{m}\right)^2\right]Y^{L}\left(\hat{\boldsymbol{S}_{m}}\right)d\hat{\boldsymbol{S}_{m}} \nonumber\\
&=&\sum_{m=1}^{n}C_{m} \frac{u_{L}\left(\boldsymbol{R},\boldsymbol{S}_{m}\right)}{\boldsymbol{R}}Y^L\left(\hat{\boldsymbol{R}}\right),
\end{eqnarray}
with
\begin{eqnarray}\label{wave7}
u_{L}(\boldsymbol{R},\boldsymbol{S}_{m})&=&\sqrt{4\pi}\left(\frac{6}{5\pi b^2}\right)^{3/4}\textbf{R}~e^{-\frac{3}{5b^2}\left(\boldsymbol{R}-\boldsymbol{S}_{m}\right)^2}  \nonumber\\
&&\times m^L j_{L}\left(-i\frac{6}{5b^2}S_{m}\right),
\end{eqnarray}
where $C_{m}$ is expansion coefficients, n is the number of the Gaussian bases, which is determined by the stability of the results, and $j_{L}$ is the Lth spherical Bessel function. Then the relative motion wave function $\chi(\textbf{R})$ is
\begin{eqnarray}\label{wave8}
\chi(\textbf{R})&=&\frac{1}{\sqrt{4\pi}}\sum_{L}\left(\frac{6}{5\pi b^2}\right)^{3/4} \nonumber\\
&&\times \sum_{m=1}^{n}C_{m}\int e^{-\frac{3}{5b^2}\left(\textbf{R}-\textbf{S}_{m}\right)^2}Y^L\left(\hat{\boldsymbol{S}_{m}}\right)d\hat{\boldsymbol{S}_{m}}.
\end{eqnarray}
After the inclusion of the center of mass motion,
\begin{eqnarray}\label{wave9}
\Phi_{G}(\textbf{R}_{G})&=&\left(\frac{5}{\pi b^2}\right)^{3/4}e^{-\frac{5}{2b^2}\textbf{R}_{G}},
 \end{eqnarray}
the total wave function Eq.(\ref{wave1}) can be rewritten as
\begin{eqnarray}\label{wave10}
\Psi_{5q}&=&\mathcal{A}\sum_{m,L}C_{i,L}\int\frac{1}{\sqrt{4\pi}}\prod_{\alpha=1}^{3}\Phi_{\alpha}(S_{m})\prod_{\beta=4}^{5}\Phi_{\beta}(-S_{m}) \nonumber\\
&&\left[\left [\eta_{I_{A}S_{A}}\eta_{I_{B}S_{B}}\right]^{IS}Y^{L}(\hat{\textbf{S}_{m}})\right]^J\left[\chi_{c}(A)\chi_{c}(B)\right]^{[\sigma]}.
\end{eqnarray}
where $\Phi_{\alpha}(S_{m})$ and $\Phi_{\beta}(-S_{m})$ are the single-particle orbital wave function with different reference centers, which specific form can be seen in Eq.~(\ref{wave0}).

With the reformulated ansatz as shown in Eq.~(\ref{wave10}), the RGM equation becomes an algebraic eigenvalue equation,
\begin{eqnarray}
  \sum_{j,L}C_{J,L}H_{i,j}^{L,L^{'}} &=& E\sum_{j}C_{j,L^{'}}N_{i,j}^{L^{'}},
\end{eqnarray}
where $N_{i,j}^{L^{'}}$ and $H_{i,j}^{L,L^{'}}$ are the  overlap of the wave functions and the matrix elements of the Hamiltonian, respectively. By solving the generalized eigenvalue problem, we can obtain the energies of the 5-quark systems $E$ and the corresponding expansion coefficient $C_{j,L}$. Finally, the relative motion wave function between two clusters can be obtained by substituting the $C_{j,L}$ into Eq.~(\ref{wave6}).

For a scattering problem, the relative wave function is expanded as
\begin{eqnarray}
  \chi_{L}\left(\textbf{R}\right) &=& \sum_{m=1}^{n}C_{m} \frac{u_{L}\left(\boldsymbol{R},\boldsymbol{S}_{m}\right)}{\boldsymbol{R}}Y^L\left(\hat{\boldsymbol{R}}\right),
\end{eqnarray}
with
\begin{equation}
\tilde{u}_{L}\left(\boldsymbol{R},\boldsymbol{S}_{m}\right)=\left \{ \begin{array}{ll}
\alpha_{m}u_{L}\left(\boldsymbol{R},\boldsymbol{S}_{m}\right), &
\boldsymbol{R} \leq \boldsymbol{R}_{C} \\
\left[h_{L}^{-}\left(\boldsymbol{k},\boldsymbol{R}\right)-\boldsymbol{S}_{m}h_{L}^{+}(\boldsymbol{k},\boldsymbol{R})\right]R_{AB}, &\boldsymbol{R} \geq \boldsymbol{R}_{C} \end{array} \right.\label{QDCSM-vc}
\end{equation}
where $u_{L}$ is formed Eq.~(\ref{wave7}), $h^{\pm}_{L}$ is the $L$th spherical Hankel functions, $k$ is the momentum of the relative motion with $k=\sqrt{2\mu E_{cm}}$, $\mu$ is the reduced mass of two hadrons of the open channel, $E_{cm}$ is the incident energy, and $R_{c}$ is a cutoff radius beyond which all the strong interaction can be disregarded. Besides, $\alpha_{m}$ and $\boldsymbol{S}_{m}$ are complex parameters which are determined by the smoothness condition at $ \boldsymbol{R}= \boldsymbol{R}_{C}$ and $C_{m}$ satisfy $\Sigma_{m=1}^{n}C_{m}=1$. After performing the variational procedure, a $L$th partial-wave equation for the scattering problem can be deduced as
\begin{eqnarray}\label{wave11}
  \sum_{j=1}^{n} \mathcal{L}^{L}_{ij}C_{j}&=& \mathcal{M}_{i}^{L}(i=0, 1,..., n-1),
\end{eqnarray}

with
\begin{eqnarray}
 \mathcal{L}^{L}_{ij}&=& \mathcal{K}_{ij}^{L}-\mathcal{K}_{i0}^{L}-\mathcal{K}_{0j}^{L}+\mathcal{K}_{00}^{L},\\
 \mathcal{M}_{i}^{L}&=& \mathcal{K}_{00}^{L}-\mathcal{K}_{i0}^{L},
\end{eqnarray}

and

\begin{eqnarray}
\mathcal{K}_{ij}^{L}&=&\left<\hat{\phi_{A}}\hat{\phi_{B}}\frac{u_{L}\left(\boldsymbol{R^{\prime}},\boldsymbol{S}_{m}\right)}{\boldsymbol{R^{\prime}}}Y^{L}\left(\boldsymbol{R^{\prime}}\right)\mid H-E \mid\right.\\
& & \left. \mathcal{A}\left[\hat{\phi_{A}}\hat{\phi_{B}}\frac{u^{L}(\boldsymbol{R},\boldsymbol{S}_{m})}{\boldsymbol{R}}Y^{L}(\boldsymbol{R})\right]   \right>.
\end{eqnarray}

By solving Eq.(\ref{wave11}), we can obtain the expansion coffefficients $C_{i}$, then the $S-$ matrix element $S_{L}$ and the phase shifts $\delta_{L}$ are given by
\begin{eqnarray}
S_{L}=e^{2i\delta_{L}}=\sum_{i=1}^{n}C_{i}s_{i}
\end{eqnarray}
Finally, the cross-section can be obtained from the scattering phase shifts by the formula:
\begin{eqnarray}\label{wave12}
\sigma_{L} =\frac{4\pi}{k^2}\cdot\left(2L+1\right)\cdot \sin^{2}\delta_{L}
\end{eqnarray}

In addition, based on Eq.~(\ref{wave12}), we can use Eq.~(\ref{wave13}) to obtain the scattering length $a_{0}$ and the effective range $r_{0}$ at the low-energy scattering phase shift, the Eq.~(\ref{wave13}) takes the form of
\begin{eqnarray}\label{wave13}
  k \cot{\delta} &=& -\frac{1}{a_{0}}+\frac{1}{2}r_{0}k^{2}+O(k^4)
\end{eqnarray}
where k represents the momentum of the relative with $k=\sqrt{2\mu E_{c.m}}$, where $\mu$ and $E_{c.m}$ are the reduced mass of two hadrons and the incident energy, respectively.

According to above results, the wave number $\alpha$ can be available by the relation~\cite{Babenko:2003js}:
\begin{eqnarray}\label{wave14}
  r_{0}&=&\frac{2}{\alpha}\left(1-\frac{1}{\alpha a_{0}} \right),
\end{eqnarray}
Finally the binding energy $B^{\prime}$ is calculated according to the relation:
\begin{eqnarray}\label{wave15}
  E_{B}^{\prime}=\frac{\hbar^{2}\alpha^{2}}{2 \mu}.
\end{eqnarray}

\end{document}